\documentclass [11pt]{article}
\usepackage {amsmath,amssymb}

\newcounter{oftheorem}[section]
\newenvironment{mytheorem}[1]%
  {\begin{trivlist}
  \renewcommand{\theoftheorem}{#1~\thesection.\arabic{oftheorem}}
  \refstepcounter{oftheorem}
  \item[\hspace{\labelsep}\bf\thesection.\arabic{oftheorem}.\,#1.]}%
  {\end{trivlist}}

\newenvironment{definition}{\begin{mytheorem}{Definition}}{\end{mytheorem}}
\newenvironment{proposition}{\begin{mytheorem}{Proposition}\sl}{\end{mytheorem}}
\newenvironment{theorem}{\begin{mytheorem}{Theorem}\sl}{\end{mytheorem}}
\newenvironment{example}{\begin{mytheorem}{Example}}{\end{mytheorem}}
\newenvironment{proof}%
  {\begin{trivlist}
  \item[\hspace{\labelsep}\bf Proof.]}%
  {\end{trivlist}}

\newcounter{cislo}
\newenvironment{conditions}%
  {\begin{list}{\textup{(\arabic{cislo})}}{\usecounter{cislo}
   \itemsep0pt\topsep5pt\parsep\parskip
   \settowidth{\labelwidth}{\upshape(OC)}\labelsep 1ex
   \leftmargin\labelwidth\addtolength{\leftmargin}{\labelsep}
   \addtolength{\leftmargin}{\parindent}}}%
  {\end{list}}

\def\0{{\bf 0}}                        % the least element
\def\1{{\bf 1}}                        % the greatest element
\def\N {\mathbb{N}}                    % natural numbers
\def\card {\mathop {\rm card}\nolimits}
\def\st {\,{:}\linebreak[0]\,\;}       % such that

\begin {document}

\title {Atomistic and orthoatomistic effect algebras}
\author {Josef Tkadlec%
\thanks {%
Department of Mathematics, Faculty of Electrical Engineering,
Czech Technical University, 166\,27 Praha, Czech Republic,
tkadlec@fel.cvut.cz.}}
\date {}

\maketitle

\begin {abstract}
We characterize atomistic effect algebras, prove that a weakly
orthocomplete Archimedean atomic effect algebra is orthoatomistic and
present an example of an orthoatomistic orthomodular poset that is not
weakly orthocomplete.
\end {abstract}

\section {Introduction}

One of the basic concepts in the foundation of quantum physics is a quantum
effect that play an important role in the theory of the so-called unsharp
measurements \cite {Busch, DP:NewTrends}. Quantum effects are studied within
a general algebraic framework called an effect algebra \cite {DP:NewTrends,
FB:Effect, GG:Toward}.

An important role in quantum structures play atoms (minimal nonzero
elements) especially if every element of the structure can be built up from
atoms, i.e., if the structure is atomistic or orthoatomistic---hence these
properties are of particular interest \cite {FB:Effect, G:AParticular,
O:Alternative, R:Orthogonal, T:Central}.

In this paper we generalize some results concerning atomistic and
orthoatomistic quantum structures and present a few illustrating examples.

\section {Basic notions and properties}

\begin {definition}
An \emph {effect algebra} is an algebraic structure $(E,\oplus,\0,\1)$ such
that $E$ is a set, $\0$ and $\1$ are different elements of $E$ and $\oplus$
is a partial binary operation on $E$ such that for every $a,b,c \in E$ the
following conditions hold:
  \begin {conditions}
  \item $a \oplus b = b \oplus a$ if $a \oplus b$ exists,
  \item $(a \oplus b) \oplus c = a \oplus (b \oplus c)$ if $(a \oplus b)
        \oplus c$ exists,
  \item there is a unique $a'\in E$ such that $a \oplus a' = \1$ (\emph
        {orthosupplement}),
  \item $a=\0$ whenever $a \oplus \1$ is defined.
  \end {conditions}
\end {definition}

For simplicity, we use the notation $E$ for an effect algebra. A partial
ordering on an effect algebra $E$ is defined by $a \le b$ iff there is a $c
\in E$ such that $b = a \oplus c$. Such an element $c$ is unique (if it
exists) and is denoted by $b \ominus a$. $\0$ ($\1$, resp.) is the least
(the greatest, resp.) element of $E$ with respect to this partial ordering.
For every $a,b \in E$, $a''=a$ and $b' \le a'$ whenever $a \le b$. It can be
shown that $a \oplus \0 = a$ for every $a \in E$ and that a \emph
{cancellation law} is valid: for every $a,b,c \in E$ with $a \oplus b \le a
\oplus c$ we have $b \le c$. An \emph {orthogonality} relation on $E$ is
defined by $a \perp b$ iff $a \oplus b$ exists (iff $a \le b'$). See,
e.g.,~\cite {DP:NewTrends,FB:Effect}.

Obviously, if $a \perp b$ and $a \lor b$ exist in an effect algebra, then
$a \lor b \le a \oplus b$. The reverse inequality need not be true (it holds
in orthomodular posets).

\begin {definition}
Let $E$ be an effect algebra. An element $a \in E$ is \emph {principal} if
$b \oplus c \le a$ for every $b,c \in E$ such that $b,c \le a$ and $b \perp
c$.
\end {definition}

\begin {definition}
An \emph {orthoalgebra} is an effect algebra $E$ in which, for every $a \in
E$, $a=\0$ whenever $a \oplus a$ is defined.

An \emph {orthomodular poset} is an effect algebra in which every element is
principal.

An \emph {orthomodular lattice} is an orthomodular poset that is a
lattice.
\end {definition}

Every orthomodular poset is an orthoalgebra. Indeed, if $a \oplus a$ is
defined then $a \oplus a \le a = a \oplus \0$ and, according to the
cancellation law, $a \le \0$ and therefore $a=\0$.

Orthomodular posets are characterized as effect algebras such that $a \oplus
b = a \lor b$ for every orthogonal pair $a,b$. (See~\cite {FB:Effect,
FGR:Filters}.) Let us remark that an orthomodular poset is usually defined
as a bounded partially ordered set with an orthocomplementation in which the
orthomodular law is valid.

\begin {definition}
Let $E$ be an effect algebra. The \emph {isotropic index} of an element $a
\in E$ is $\sup \{n\in\N \st na \ \text {is defined}\}$, where $na =
\bigoplus_{i=1}^n a$ is the sum of $n$ copies of $a$.

An effect algebra is \emph {Archimedean} if every its nonzero element has a
finite isotropic index.
\end {definition}

The isotropic index of $\0$ is $\infty$. In an orthoalgebra we have that $a
\oplus a$ is defined only for $a=\0$, hence the isotropic index of every
nonzero element is $1$. Therefore we obtain:

\begin {proposition}
Every orthoalgebra is Archimedean.
\end {proposition}

\begin {definition}
Let $E$ be an effect algebra. A system $(a_i)_{i\in I}$ of (not necessarily
distinct) elements of $E$ is called \emph {orthogonal}, if $\bigoplus_{i\in
F} a_i$ is defined for every finite set $F \subset I$. We define
$\bigoplus_{i \in I} a_i = \bigvee \{\bigoplus_{i\in F} a_i \st F \subset I
\ \text {is finite}\}$ if the supremum exists.

An effect algebra $E$ is \emph {orthocomplete} if $\bigoplus_{i\in I} a_i$
is defined for every orthogonal system $(a_i)_{i\in I}$ of elements of $E$.

An effect algebra $E$ is \emph {weakly orthocomplete} if for every
orthogonal system $(a_i)_{i\in I}$ of elements of $E$ either
$\bigoplus_{i\in I} a_i$ exists or there is no minimal upper bound of the
set $\{\bigoplus_{i \in F} a_i \st F \subset I \ \text{is finite}\}$ in $E$.
\end {definition}

Every pair of elements of an orthogonal system is orthogonal. On the other
hand, there are mutually orthogonal elements that do not form an orthogonal
system if the effect algebra is not an orthomodular poset. Since only the
zero element is orthogonal to itself in an orthoalgebra, we may consider
sets instead of systems in orthoalgebras.

\begin {proposition}
Every orthocomplete effect algebra is Archimedean.
\end {proposition}

\begin {proof}
Let $E$ be an orthocomplete effect algebra and let $a \in E$ has an infinite
isotropic index. There is an element $b \in E$ such that $b =
\bigoplus_{n\in\N} a = \bigvee_{n\in\N} na$. Since $a \le b$, there is an
element $c \in E$ such that $b = a \oplus c$. For every $n\in\N$ we have $a
\oplus c = b \ge (n+1) a = a \oplus na$ and therefore, according to the
cancellation law, $c \ge na$. Hence $ c \oplus \0 = c \ge \bigvee_{n\in\N}
na = b = c \oplus a$ and, according to the cancellation law, $\0 \ge a$ and
therefore $a=\0$.
\end {proof}

\begin {definition}
An \emph {atom} of an effect algebra $E$ is a minimal element of $E
\setminus \{\0\}$.

An effect algebra is \emph {atomic} if every nonzero element dominates an
atom (i.e., there is an atom less than or equal to it).

An effect algebra is \emph {atomistic} if every nonzero element is a
supremum of a set of atoms (i.e., of the set of all atoms it dominates).

An effect algebra is \emph {orthoatomistic} if every nonzero element is a
sum of a set of atoms.
\end {definition}

It is easy to see that every atomistic and every orthoatomistic effect
algebra is atomic and that every orthoatomistic orthomodular poset is
atomistic. There are atomic orthomodular posets that are not atomistic~\cite
{G:AParticular}, atomistic orthomodular posets that are not
orthoatomistic~\cite {O:Alternative} and orthoatomistic orthoalgebras that
are not atomistic---e.g., the so-called Wright triangle~\cite
[Example~2.13]{FGR:Filters}.

\section {Results}

First, let us present a characterization of atomistic effect algebras that
generalizes the result of~\cite {O:Alternative} stated for orthomodular posets.

\begin {definition}
An effect algebra $E$ is \emph {disjunctive} if for every $a,b \in E$ with
$a \not\le b$ there is a nonzero element $c \in E$ such that $c \le a$ and
$c \land b = \0$.
\end {definition}

\begin {theorem}
An effect algebra is atomistic if and only if it is atomic and disjunctive.
\end {theorem}

\begin {proof}
Let $E$ be an effect algebra and let us for every $x \in E$ denote by $A_x$
the set of atoms dominated by $x$.

$\Rightarrow$: Obviously, every atomistic effect algebra is atomic. Let $a,b
\in E$ such that $a \not\le b$. Then there is an atom $c \in A_a \setminus
A_b$, hence $c \le a$ and $c \land b = \0$.

$\Leftarrow$: Let us prove that $a \le b$ for every nonzero $a \in E$ and for
every upper bound $b \in E$ of $A_a$ (hence $a = \bigvee A_a$). Let us
suppose that $a \not\le b$ and seek a contradiction. Since $E$ is
disjunctive, there is a nonzero element $c \in E$ such that $c \le a$ and $c
\land b = \0$. Since $E$ is atomic, there is an atom $d \in E$ such that $d
\le c$. Hence $d \le a$ and $d \land b = \0$. Since $d$ is an atom, $d
\not\le b$ and therefore $d \in A_a \setminus A_b$---a contradiction.
\end {proof}

Before stating the second main result of this paper, let us discuss
relations of some properties.

\begin {proposition}
Let $E$ be an effect algebra fulfilling at least one of the following
conditions:
  \begin {conditions}
  \item [\rm(OC)] $E$ is orthocomplete.
  \item [\rm(L)] $E$ is a lattice.
  \end {conditions}
Then $E$ is weakly orthocomplete.
\end {proposition}

\begin {proof}
(OC): Obvious.

(L): Let $(a_i)_{i\in I}$ be an orthogonal system of elements of $E$. Let us
show that if a minimal upper bound $a$ of the set $A = \{\bigoplus_{i\in F}
a_i \st F \subset I\ \text {is finite}\}$ exists then $a = \bigvee A$. Let
$b$ be an upper bound of $A$. Then $b \land a \le a$ is an upper bound of
$A$ and, since $a$ is minimal, $b \land a = a$. Hence $a \le b$.
\end {proof}

Let us present examples showing that the scheme of implications in the
previous proposition cannot be improved.

\begin {example}
Let $X$ be a countable infinite set. Let $E$ be a family of finite and
cofinite subsets of $X$ with the $\oplus$ operation defined as the union of
disjoint sets. Then $(E,\oplus,\emptyset,X)$ is an orthomodular lattice (it
forms a Boolean algebra) that is not orthocomplete.
\end {example}

\begin {example}
Let $X$ be a 6-element set. Let $E$ be the family of even-element subsets of
$X$ with the $\oplus$ operation defined as the union of disjoint sets from
$E$. Then $(E,\oplus,\emptyset,X)$ is a finite (hence orthocomplete)
orthomodular poset that is not a lattice.
\end {example}

\begin {example}
Let $X_1,X_2,X_3,X_4$ be mutually disjoint infinite sets,
$X = \bigcup_{i=1}^4 X_i$,
  \begin {align*}
  E_0 &= \{\emptyset, X_1 \cup X_2, X_2 \cup X_3, X_3 \cup X_4, X_4 \cup X_1, X\}\,,\\
  E   &= \{(A \setminus F) \cup (F \setminus A) \st F \subset X \ \text {is
            finite},\ A \in E_0 \}\,,
  \end {align*}
$A \oplus B = A \cup B$ for disjoint $A,B \in E$. Then $(E, \oplus,
\emptyset, X)$ is a weakly orthocomplete orthomodular poset that is neither
orthocomplete (e.g., $\bigvee \bigl\{ \{x\} \st x \in X_1 \bigr\}$ does not
exist) nor a lattice (e.g., $(X_1 \cup X_2) \land (X_2 \cup X_3)$ does not
exist).
\end {example}

\begin {theorem} \label {T:orthoatomistic}
Every weakly orthocomplete Archimedean atomic effect algebra is
orthoatomistic.
\end {theorem}

\begin {proof}
Let $E$ be a weakly orthocomplete Archimedean atomic effect algebra and let
$a \in E \setminus \{\0\}$. Let us consider orthogonal systems of atoms such
that their finite sums are dominated by~$a$. Since $E$ is atomic, there are
such systems. Since $E$ is Archimedean, using the Zorn's lemma we obtain
that there is a maximal such system, $M$. Let us show that $a$ is a minimal
upper bound of the set $A = \{ \bigoplus F \st F \subset M \ \text {is
finite}\}$. Indeed, if there is an upper bound $b \in E$ of $A$ such that
$b<a$ then $a \ominus b \neq \0$, there is an atom $c \in E$ such that $c
\le a \ominus b$ and therefore $c \oplus \bigoplus F \le a$ for every finite
set $F \subset M$---this contradicts to the maximality of $M$. Since $E$ is
weakly orthocomplete, $a = \bigvee A = \bigoplus M$.
\end {proof}

The previous theorem generalizes the result of~\cite {O:Alternative} stated
for weakly orthocomplete atomic orthomodular posets, the result of~\cite
[Proposition~4.11]{FB:Effect} stated for chain finite effect algebras and
the result of~\cite [Theorem~3.1]{R:Orthogonal} stated for lattice
Archimedean atomic effect algebras.

None of the assumptions in \ref {T:orthoatomistic} can be omitted. Indeed,
there are atomistic orthomodular posets that are not orthoatomistic~\cite
{O:Alternative}, Boolean algebras that are not atomic (e.g., $\exp
\N|_{F(\N)}$ where $F(\N)$ denotes the family of finite subsets of the set
$\N$ of natural numbers), and, as the following example shows, weakly
orthocomplete atomic effect algebras that are not orthoatomistic.

\begin {example}
Let $E = \{0, 1, 2, \dots, n, \dots, n', \dots, 2', 1', 0'\}$ with the
$\oplus$ operation defined by $m \oplus n = m + n$ for every $m,n \in \N$
and $m \oplus n' = (n-m)'$ for every $m,n \in \N$ with $m \le n$. Then $(E,
\oplus, 0, 0')$ is an atomic effect algebra (it forms a chain) that is
weakly orthocomplete. Indeed, if an orthogonal system $M$ of nonzero
elements of $E$ is finite then $\bigoplus M$ is defined; if $M$ is infinite
then the set of finite sums of elements of $M$ forms an unbounded set of
natural numbers and therefore does not have a minimal upper bound. The
effect algebra is not orthoatomistic because no element $n'$, $n \in \N$, is
a sum of atoms.
\end {example}

Let us present an example that an orthoatomistic orthomodular poset need not
be weakly orthocomplete.

\begin {example}
Let $X, Y$ be disjoint infinite countable sets,
  \begin {align*}
  E_0 &= \{A \subset (X \cup Y) \st \card (A \cap X) = \card (A \cap Y) \
           \text {is finite}\}\,,\\
  E   &= E_0 \cup \{(X \cup Y) \setminus A \st A \in E_0 \}\,,
  \end {align*}
$A \oplus B = A \cup B$ for disjoint $A,B \in E$. Then $(E, \oplus,
\emptyset, X \cup Y)$ is an orthomodular poset. It is orthoatomistic because
for every nonempty $A \in E$ we have $\card (A \cap X) = \card (A \cap Y)$,
there is a bijection $f \st (A \cap X) \to (A \cap Y)$ and $A = \bigoplus
\bigl\{ \{x, f(x)\} \st x \in (A \cap X) \bigr\}$. The orthomodular poset is
not weakly orthocomplete because for $x_0 \in X$, $y_0 \in Y$ there is a
bijection $f \st X \to (Y \setminus \{y_0\})$ and the orthogonal set $\bigl\{
\{x,f(x)\} \st x \in X \setminus \{x_0\} \bigr\}$ has different minimal
upper bounds $(X \cup Y) \setminus \{x_0, f(x_0)\}$ and $(X \cup Y)
\setminus \{x_0, y_0\}$.
\end {example}

\section *{Acknowledgements}

The work was supported by the grant of the Grant Agency of the Czech
Republic no.~201/07/1051 and by the research plan of the Ministry of
Education of the Czech Republic no.~6840770010.


\begin {thebibliography}{10}

\bibitem {Busch} Busch, P.:
\emph {The Quantum Theory of Measurements}.
Springer-Verlag, New York, 2002.

\bibitem {DP:NewTrends} Dvure\v{c}enskij, A., Pulmannov\'a, S.:
\emph {New Trends in Quantum Structures}.
Kluwer Academic Publishers, Bratislava, 2000.

\bibitem {FB:Effect} Foulis, D.~J., Bennett, M.~K.:
\emph {Effect algebras and unsharp quantum logics},
Found. Phys. {\bf 24} (1994), 1331--1352.

\bibitem {FGR:Filters} Foulis, D., Greechie, R., R\"uttimann, G.:
\emph {Filters and supports in orthoalgebras}.
Internat. J. Theoret. Phys. {\bf 31} (1992), 789--807.

\bibitem {GG:Toward} Giuntini, R., Greuling, H.:
\emph {Toward a formal language for unsharp properties}.
Found. Phys. {\bf 19} (1989), 931--945.

\bibitem {G:Disjunctivity} Godowski, R.:
\emph {Disjunctivity and orthodisjunctivity in orthomodular posets}.
Demonstratio Math. {\bf 12} (1979), 1043--1049.

\bibitem {G:AParticular} Greechie, R. J.:
\emph {A particular non-atomistic orthomodular poset}.
Comm. Math. Phys. {\bf 14} (1969), 326--328.

\bibitem {O:Alternative} Ovchinnikov, P. G.:
\emph {On alternative orthomodular posets},
Demonstratio Math. {\bf 27} (1994), 89--93.

\bibitem {R:Orthogonal} Rie\v{c}anov\'a, Z.:
\emph {Orthogonal sets in effect algebras},
Demonstratio Math. {\bf 34} (2001), 525--532.

\bibitem {T:Central} Tkadlec, J.:
\emph {Central elements of atomic effect algebras}.
Internat. J. Theoret. Phys. {\bf 44} (2005), 2257--2263.

\end {thebibliography}
\end {document}